\documentclass[10pt,preprint]{aastex}

\received{2002 October 4}
\begin{document}

\title{{\it Chandra} Observation of the X-Ray Source Population of NGC 6946}

\author{S. S. Holt\altaffilmark{1}, E. M. Schlegel\altaffilmark{2},
U. Hwang\altaffilmark{3}, and R. Petre\altaffilmark{3}}

\altaffiltext{1}{Olin College, Needham, MA}
\altaffiltext{2}{Center for Astrophysics, SAO/HCO, Cambridge, MA}
\altaffiltext{3}{Laboratory for High Energy Astrophysics, GSFC, Greenbelt, MD}

\begin{abstract}

We present the results of a study of discrete X-ray sources in NGC
6946 using a deep {\it Chandra} ACIS observation.  Based on the slope
of the log N-log S distribution and the general correlation of sources
with the spiral arms, we infer that the overall discrete source sample 
in NGC 6946 is dominated by high mass X-ray binaries, in contrast to the 
source distributions in M31 and the Milky Way.  This is consistent with 
the higher star formation rate in NGC 6946 than in those galaxies.  
We find that the strong X-ray sources in the region of the galactic center 
do not correlate in detail with images of the region in the near-IR, 
although one of them may be coincident with the galactic center.  
The non-central ultra-luminous X-ray source in NGC 6946, previously 
identified with a supernova remnant, has an X-ray spectrum and luminosity 
that is inconsistent with either a traditional pulsar wind nebula or 
a blast wave remnant.

\end{abstract}


\section{Introduction}

NGC 6946 is a relatively nearby grand-design spiral galaxy.  It has
been classified as having strong nuclear starburst activity
\citep{Elm98}, and the six historical supernova remnants recorded
during the past century attest to a high rate of star formation in its
spiral arms \citep{SGC98}.  Its face-on aspect provides the
opportunity to observe the total X-ray source population of a spiral
galaxy and to compare its characteristics with those of other
nearby spiral galaxies.

The distance to NGC 6946 is estimated between 5.1 Mpc \citep{deV79} and 
5.9 Mpc \citep{KSH00}.  We adopt 5.9 Mpc (see \citealt{Lar02}), so that 
our luminosity estimates would be reduced by one-quarter for a distance 
of 5.1 Mpc.  The low galactic inclination of NGC 6946 indicates that the
preponderance of photoelectric absorption in its direction arises from
gas in the Milky Way.  From the extinction maps of \citep{SFD98}, the
average hydrogen column density in the direction of NGC 6946 should be
N$_H$ = 3-5$\times$10$^{21}$ cm$^{-2}$.

Several recent studies of NGC 6946 have been devoted to its X-ray
source, radio source and supernova remnant populations.  \cite{Sch94}
and \cite{SBF00} detected a total of fourteen X-ray sources with
{\it Einstein}, {\it Rosat} and {\it ASCA}, four of which were 
coincident with entries in the 118-member radio source catalog of 
\cite{LDG97}(hereafter LDG).  \cite{LD01} invoked the simple 
requirement of a nonthermal radio spectrum to identify 35 members of 
the LDG catalog that can be considered
``middle-aged'' (i.e., $\sim$10$^4$ years) supernova remnants.  Of
these, there is only one match with the 27-member optically selected
SNR catalog of \cite{MF97}; this single coincidence (hereafter MF16)
was further identified with the strongest X-ray source in NGC 6946
detected by \cite{SBF00}.

In this paper we use X-ray imaging and spectral data from the {\it Chandra}
observatory to study the X-ray source population of this face-on
spiral galaxy in terms of its logN-logS distribution, spatial
distribution, and spectral distribution.  We compare these properties
to those of other spiral galaxies with varying star formation rates.

\section{Source Detection with {\it Chandra}}

\begin{figure*}
\begin{center}
\scalebox{0.5}{\rotatebox{-90}{\includegraphics{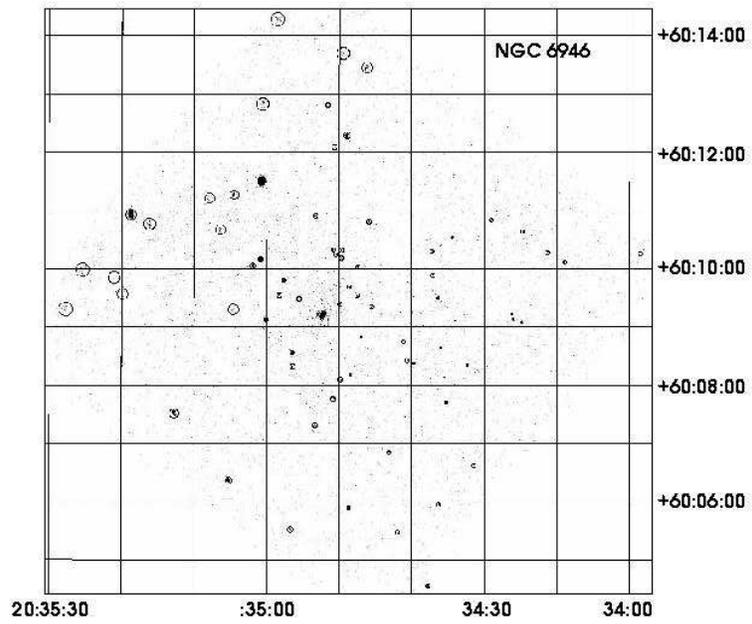}}}
\caption{60 ks {\it Chandra} observation of NGC 6946 on S3 ACIS chip (8 arc
min on a side).  Photons with energy-equivalent pulse height $<$ 0.3 keV
or $>$ 5 keV are excluded.}
\label{raw_fig}
\end{center}
\end{figure*}

NGC 6946 was observed by the ACIS instrument aboard the {\it Chandra
Observatory} for approximately 60 ks on September 7, 2001.  The 8$'$ x
8$'$ field-of-view of a single CCD chip allows almost all of the
galaxy to be imaged on the back-side-illuminated S3 chip.  The raw
data, with photons of pulse-height equivalent to energies below 0.3
keV or above 5 keV removed, are exhibited in Figure~\ref{raw_fig}.

Source identification was done primarily with standard tools in the
CIAO ({\it Chandra} Interactive Analysis of Observations) toolbox.
{\tt Celldetect} is a simple boxcar filter algorithm in the CIAO
software package that utilizes a variable detection cell size (i.e.,
increasing with angular distance from the telescope aim point to match
the telescope point spread function, PSF) that would include at least
80\% of the counts from a point source.  We applied the Celldetect 
algorithm to the 0.3-5 keV data of Figure~\ref{raw_fig}, for which 
the algorithm detected 46 sources at a significance of $>$3$\sigma$.  
We separately applied the same
algorithm to the data in several energy bands in order to perform some
straightforward comparative diagnostics.  There is a natural break in
the X-ray response at 2 keV, which lends itself to defining the 2-5
keV band as hard (H).  We have divided the softer X-rays into 0.3-0.5
(VS), 0.5-1 keV (S) and 1-2 keV (M). Sources that are detected in the
VS band should be foreground sources since the X-ray optical depth
through the Milky Way at 0.5 keV is $\sim$3.

We also utilized the {\tt Vtpdetect} (Voronoi Tesselation and
Percolation) algorithm, which considers asymmetric pixelation and can
be more sensitive than Celldetect for point sources at the
detection limit in uncrowded fields.  Since 1$''$ corresponds to 30 pc 
at a distance of 5.9 Mpc, for this study we are more interested in the 
sensitivity of Vtpdetect than in the imaginative way that it might
identify extended sources that would be tens or hundreds of pc in 
size.  Vtpdetect identifies 86 sources at $>$3$\sigma$ when 
applied to the 0.3-5 keV FITS file of Figure~\ref{raw_fig}.  
Visual inspection of the image results in 72 relatively firm
identifications (at least 10 source counts).  The typical background
expectation in the southwestern portion of the image is less than 1
count, but in the northeast (where the point spread function is wider
and there may be some diffuse emission) there may be as many as 3
background counts expected.  In a separate study, \cite{SHP02} are
considering the extent to which this background might be diffuse
emission in the galaxy associated with the hot phase of its
interstellar medium.  Other sources identified visually at slightly
lower statistical confidence have not been included in
Table~\ref{detsrcs} if they were not first objectively identified by
Celldetect or Vtpdetect.

\section{Characteristics of the Source Population}

As an imaging spectrometer, the ACIS detector offers the opportunity
to investigate the spatial, spectral and luminosity characteristics 
of the galactic source population.

\begin{figure*}
\begin{center}
\scalebox{0.5}{\rotatebox{-90}{\includegraphics{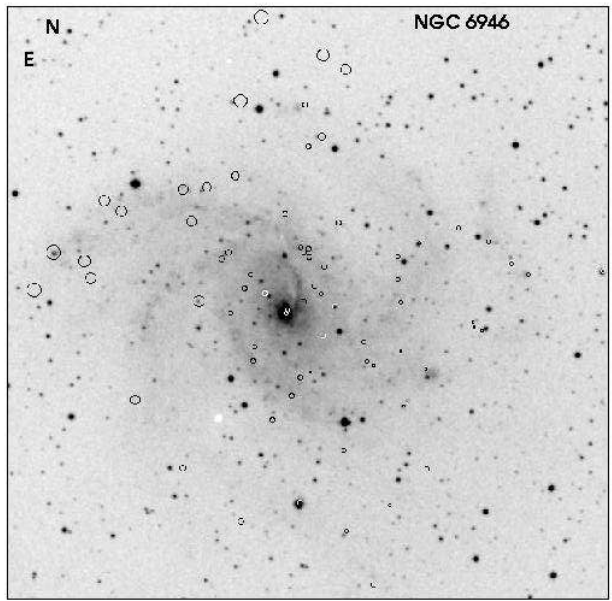}}}
\caption{The 72 X-ray source positions displayed over the optical
image of NGC 6946.  Positional error circles increase with distance
from the telescope aim point, which is approximately 3 arc minutes to
the southwest of the galactic nucleus.  The image is 8 arc min on a side.}
\label{opt_fig}
\end{center}
\end{figure*}

Although we have not performed a quantitative correlation analysis,
visual inspection of the location of the X-ray sources projected on
the optical image of NGC 6946 indicates a clear qualitative
correlation with the spiral arms (Figure~\ref{opt_fig}).  A similar
association was noted by \cite{LD01} for those radio sources that they
designated to be ``middle-aged SNR'' (sources selected on the basis of
their radio properties to be likely candidates for SNR with ages of
$\sim$10$^4$ years that are in the adiabatic phases of their
expansion); 6 of the 35 sources in their sample are coincident (within
2$''$) with one of the 72 X-ray sources catalogued here.  It should be 
noted that since the radio survey selected sources with nonthermal 
spectra (thermal radio SNRs are indistinguishable from H~II regions), 
pulsar wind nebula SNRs with 
thermal radio spectra would not have been included in that survey.
Interestingly, 8 of the 83 radio sources that do not meet the
middle-aged SNR criteria were also coincident with X-ray sources.  
In particular, source 82 in the LDG radio catalog is coincident with 
the young remnant of SN1968D and is also
identified with an X-ray source of 0.5-5 keV luminosity
$\sim$3$\times$10$^{37}$ (d$_{\rm 5.9 Mpc}$)$^2$ erg s$^{-1}$.  In
contrast to the distribution of the middle-aged radio SNR, the 27
optically-selected SNR of \cite{MF97} were not particularly
well-correlated with the young stellar population of the spiral arms;
here MF16 is the only member of the sample that has an X-ray
counterpart.

The results of fits to the spectra of all those sources for which we
have $>$300 counts are displayed in Table 2.  All of these have
inferred luminosities comparable to, or in excess, of the Eddington
limit for a 1.4 M$_{\odot}$ neutron star (L$_{\rm Edd}$ =
4$\pi$cGm$_p$M/$\sigma$$_T$= 1.3$\times$10$^{38}$ M/M$_{\odot}$ erg
s$^{-1}$).  Since the number of background counts accumulated in each
spectrum is typically $<$10, we have not subtracted background in
determining these fits.  None of the spectra have the obvious bright
K-lines that are associated with the ejecta of young (100-1000 year
old) supernova remnants in the Milky Way, although some of the spectra
can be described with optically thin thermal models that include less
prominent line emission.

Of the 32 sources with $>$40 counts ($>$2.8$\times$10$^{37}$ (d$_{\rm
5.9 Mpc}$)$^2$ erg s$^{-1}$), the general population appears to
exhibit featureless ``Crab-like'' spectra.  Comparing the S-M-H counts
from each source with those expected from various power law trial
spectra, half (16) are best characterized by 1.5$< {\Gamma} <$2.5,
with the other half evenly distributed between softer (7 with
${\Gamma}>$2.5) and harder (9 with ${\Gamma}<$1.5) spectra.  The very
brightest spectra seem, on average, to be slightly softer (see Table
2).

For source spectra that are not very steep (i.e., photon power law
index $\Gamma$ $\leq$2 or energy power law index $\alpha$ $\leq$1, or
models that have characteristic temperatures $\geq$ few tenths of a
keV), column densities in the range N$_{\rm H}$ = 3-5$\times$10$^{21}$
cm$^{-2}$ typically reduce the source flux in the 0.5-5 keV band by
about a factor of 2 relative to that which would be detected with
N$_{\rm H}$ = 0.  Using this prescription for typical source spectra,
the minimum 0.5-5 keV source luminosity at 5.9 Mpc (corresponding to
10 counts) is approximately 7$\times$10$^{36}$ (d$_{\rm 5.9 Mpc}$)$^2$
erg s$^{-1}$, and the survey should be complete down to approximately
10$^{37}$ (d$_{\rm 5.9 Mpc}$)$^2$ erg s$^{-1}$.  More than a dozen
sources have 0.5-5 keV luminosities comparable with, or in excess of,
the Eddington luminosity for a 1.4 M$_{\odot}$ neutron star.  Using
the nearer distance to NGC 6946 would not substantially change this
observation.

\begin{figure*}
\begin{center}
\scalebox{0.4}{\includegraphics{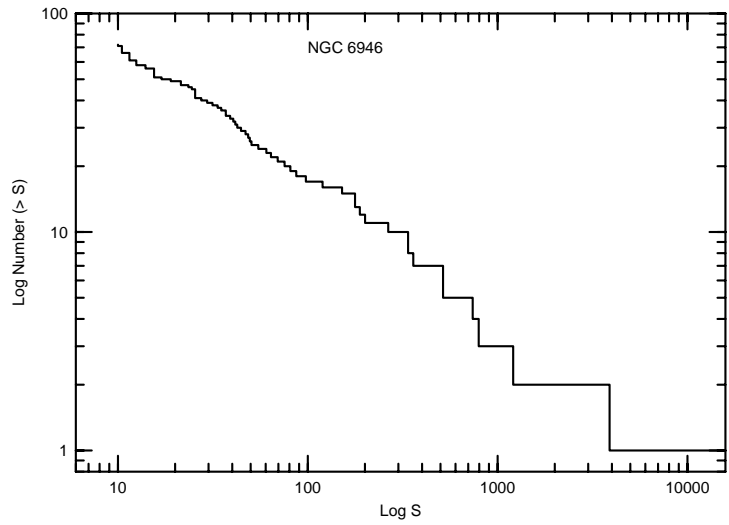}}
\caption{The integral logN/logs distribution of the 72 X-ray sources
reported from this observation of NGC 6946.  The horizontal scale has
units of 7$\times$10$^{35}$ (d$_{5.9Mpc}$)$^2$ erg s$^{-1}$.}
\label{lum_dist}
\end{center}
\end{figure*}

Figure~\ref{lum_dist} displays the traditional logN-logS distribution
of the integral number of sources N($>$S) with apparent luminosity
greater than S.  The unbinned distribution of the 72 sources detected
with luminosities $>$7$\times$10$^{36}$ (d$_{\rm 5.9 Mpc}$)$^2$ erg
s$^{-1}$ is well-fit with a power law index of 0.64$\pm$0.02 (95\%
confidence).  If we confine the fit to luminosities $>$10$^{37}$
(d$_{\rm 5.9 Mpc}$)$^2$ erg s$^{-1}$, where we expect that the survey
is complete, the fit steepens slightly to 0.68$\pm$0.03.  The
logN-logS distribution of the 20 sources with luminosities
$>$7$\times$10$^{36}$ (d$_{\rm 5.9 Mpc}$)$^2$ erg s$^{-1}$ that are
located within 2$'$ of the galactic center exhibit a somewhat flatter
distribution, with an index 0.51$\pm$0.05.  None of the sources in
Table~\ref{detsrcs} appear to have either very soft (stellar-like)
spectra or column densities much less than N$_{\rm H}$ =
3$\times$10$^{21}$ cm$^{-2}$, so that all can either be associated
with NGC 6946 or are background galaxies or AGN.  Using the 0.7-2 keV
logN-logS relationship of \cite{Ueda99}, which nicely overlaps the
region of sensitivity for the sources that we detect in NGC 6946, we
expect that the source list of Table~\ref{detsrcs} should be
contaminated with approximately 2 background sources.  These 
contaminating galaxies would typically have the relatively hard 
spectra characteristic of AGN, but young pulsar-driven supernova 
remnants or X-ray binaries in NGC 6946 could have similarly hard 
spectra.

\section{The Ultraluminous Source in NGC 6946}

The X-ray source coincident with MF16 is the brightest X-ray source
that we observe in NGC 6946, with 0.5-5 keV luminosity that we
estimate to be approximately 8$\times$10$^{39}$ (d$_{\rm 5.9
Mpc}$)$^2$ erg s$^{-1}$, based upon the two-component bremsstrahlung
fit that is consistent with the average column density in the
direction of NGC 6946.  This is in good agreement with the luminosity
estimated by \cite{Sch94} from data taken with {\it ROSAT}, but less
than that estimated by \cite{SBF00} from data taken with {\it ASCA} 
(where the higher luminosity estimate might have arisen from 
over-correcting for the broad {\it ASCA} point spread function).
Without clear evidence for a secular decrease in the MF16 luminosity,
we shall assume that all three results are consistent with a constant
source luminosity over ten years.

\cite{SBF00} have suggested that the morphology of MF16 region 
revealed by the {\it Hubble Space Telescope} 
\citep{BFS01} could imply that the X-ray emission
arises from the collision of shock fronts associated with multiple
supernovae.  The feature upon which this supposition rests is
approximately 1$''$ in angular extent, which implies a blast wave 
cavity tens of pc in size.  To traverse this distance, a blast wave 
of velocity 10$^4$ km s$^{-1}$ would require $>$10$^4$ years, an
age at which SNR blast waves are generally well past their
free-expansion phases, and well past the times at which they would 
have exhibited their maximum luminosities.  The 0.5-5 keV luminosity 
of MF16 is almost three orders of magnitude more than that of the 
brightest young blast wave remnants like Cas A.

It might be possible to explain the high luminosity in terms of a
transient phase in the cooling, where the SN shock wave is depositing
its kinetic energy into a dense ambient medium, since the emission
measure is proportional to the square of the density.  Such a scenario
for this source has been suggested by \cite{Dun00}, who argue that the
characteristics of the source can be explained by supernova ejecta
expanding into a dense nitrogen-rich circumstellar nebula created by
the high-mass progenitor.  However, as noted by the same authors, this
cannot be the complete picture, since the spectrum measured by the 
{\it ROSAT} proportional counter requires a non-thermal contribution, 
perhaps from a pulsar or a pulsar wind nebula (or a superposition of 
more than one PWN).  Moreover, the optically inferred blast wave 
velocity is less than 1000 km/s.  With the current 
{\it Chandra} data, we see that the spectrum of MF16 is
indisputably deficient in the line emission expected from an
interaction with a dense CSM, although such a component could make a
minor contribution to the overall spectrum.

\begin{figure*}
\begin{center}
\scalebox{0.4}{\rotatebox{-90}{\includegraphics{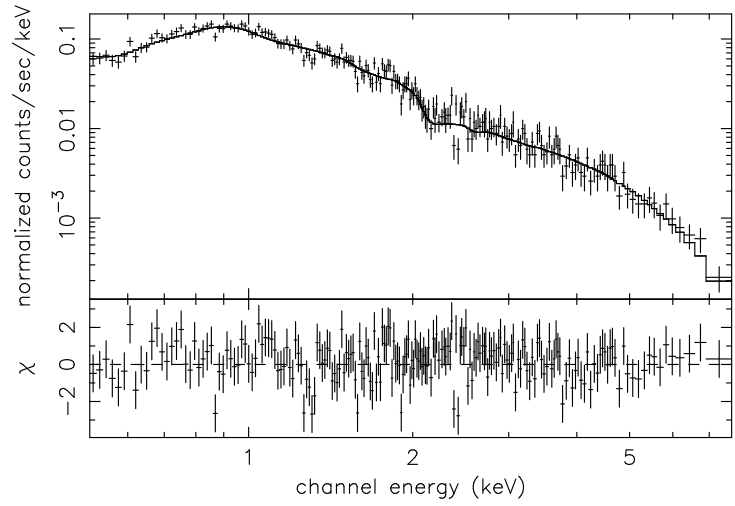}}}
\caption{The X-ray spectrum of MF16.  The model is a thermal plasma
(MEKAL) and a power law, with column density only half of the average
in the direction of NGC6946.  A two-bremss model with higher column
density is virtually indistinguishable. The fits are described in
Table~2.}
\label{mf16_spec}
\end{center}
\end{figure*}

The X-ray spectrum of MF16 has excellent statistics, containing
several thousand counts.  The lack of pronounced spectral features
(such as the dominant K-lines that are characteristic of young
blast-wave remnants) do not demand a unique model, and the spectrum
can be fit equally well with a variety of models that sum two or more
continuum components.  One example is shown in Figure~\ref{mf16_spec}.  This
composite of a power law and thermal plasma (MEKAL in the jargon of
the XSPEC fitting routine that was used here) might be consistent with
a pulsar wind nebula and the interaction of the blast wave with a
dense circumstellar shell, but both components are associated with
unusually high luminosities that are not easily explained.  Moreover,
it is not the only model that gives an acceptable fit, and the
two-component bremsstrahlung model has a higher column density that
might better represent the unabsorbed continuum.  The problem with the
bremsstrahlung model is that it requires a very low ionization age to
suppress the line emission, which is not expected for a remnant that
has been interacting with a dense circumstellar medium for at least 
the 10 years over which observations of this source have been made.

We find no evidence for the companion source that was reported by
\cite{SBF00} at a level of 7\% that of MF16 and at an angular distance
30$''$ from it .  With no indication of a source at this position, its
current level of emission must be at least an order of magnitude
weaker than when detected by Schlegel, Blair, and Fesen, indicating
that the missing source is likely to be a variable X-ray binary.

\section{Galactic Nucleus}

In the region of the galactic center, there are no sources coincident
with the J2000 nuclear position RA=308.7201, DEC=60.1538 of
\cite{Cot99}.  \cite{Car90} place the dynamical center some 5$''$ west
of the Cotton, Condon, and Arbizzani position, coincident with the K$'$
peak reported by \cite{RV95}.  Two strong X-ray sources in the
galactic center region are detected by the algorithms described in
section 2, with the stronger source located $\sim$2$''$ south of the
weaker.  A third, much weaker source 11$''$ to the east is not
detected by the algorithms (and not included in Table~\ref{detsrcs}),
although a fourth, yet weaker source to its south is detected, being
far enough from the bright central region to be unaffected by the
response of the telescope to other sources.  The bright source lies
$\sim$2.7$''$ NE (position angle $\sim$60 degrees E from N) from the
dynamical center of Carignan et al. and $\sim$8.3$''$ W from the
Cotton et al. center.

Both strong sources have spectra that are well-fit with power laws of
energy index $\Gamma$ $\sim$1.4 with the same column density.  This
suggests the possibility that there may remain some contamination of
our spectrum of the smaller by the larger, in spite of the fact that
the two sources appear to be almost resolved; it could also
indicate that the two sources have similar origins in the star-forming
nucleus of the galaxy.  Interestingly, the combination of the two
sources appears very much like the morphology observed in both the CO
image of \cite{RV95} and the H alpha image of \cite{MF97} (see
Fig.~\ref{nuc_fig_exp}).  The CO ``molecular bar'' is a roughly
rectangular structure of dimensions approximately 60$''$ by 12$''$
including the central K$'$ (and H alpha) peak of Fig.~\ref{nuc_fig_exp}
at a position angle of $\sim$135$^{\circ}$ E of N.

\begin{figure*}
\begin{center}
\scalebox{0.4}{\includegraphics{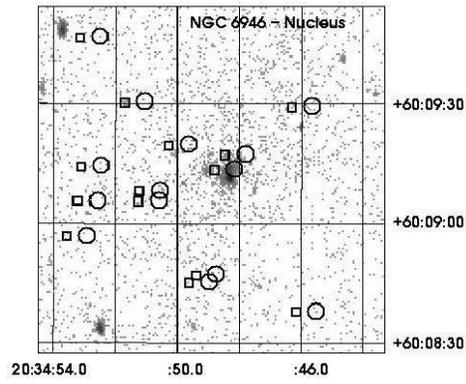}}
\caption{The inner region of NGC 6946.  This region contains at two
strong point sources, and several weaker ones.  The symbols represent
the nuclear H II regions described in \cite{Elm98} but using two
different definitions for the absolute center.  The squares are
referred to the center of \cite{Cot99}; the circles are referred to
the center of \cite{Car90}.}
\label{nuc_fig}
\end{center}
\end{figure*}

\begin{figure*}
\begin{center}
\scalebox{0.4}{\rotatebox{-90}{\includegraphics{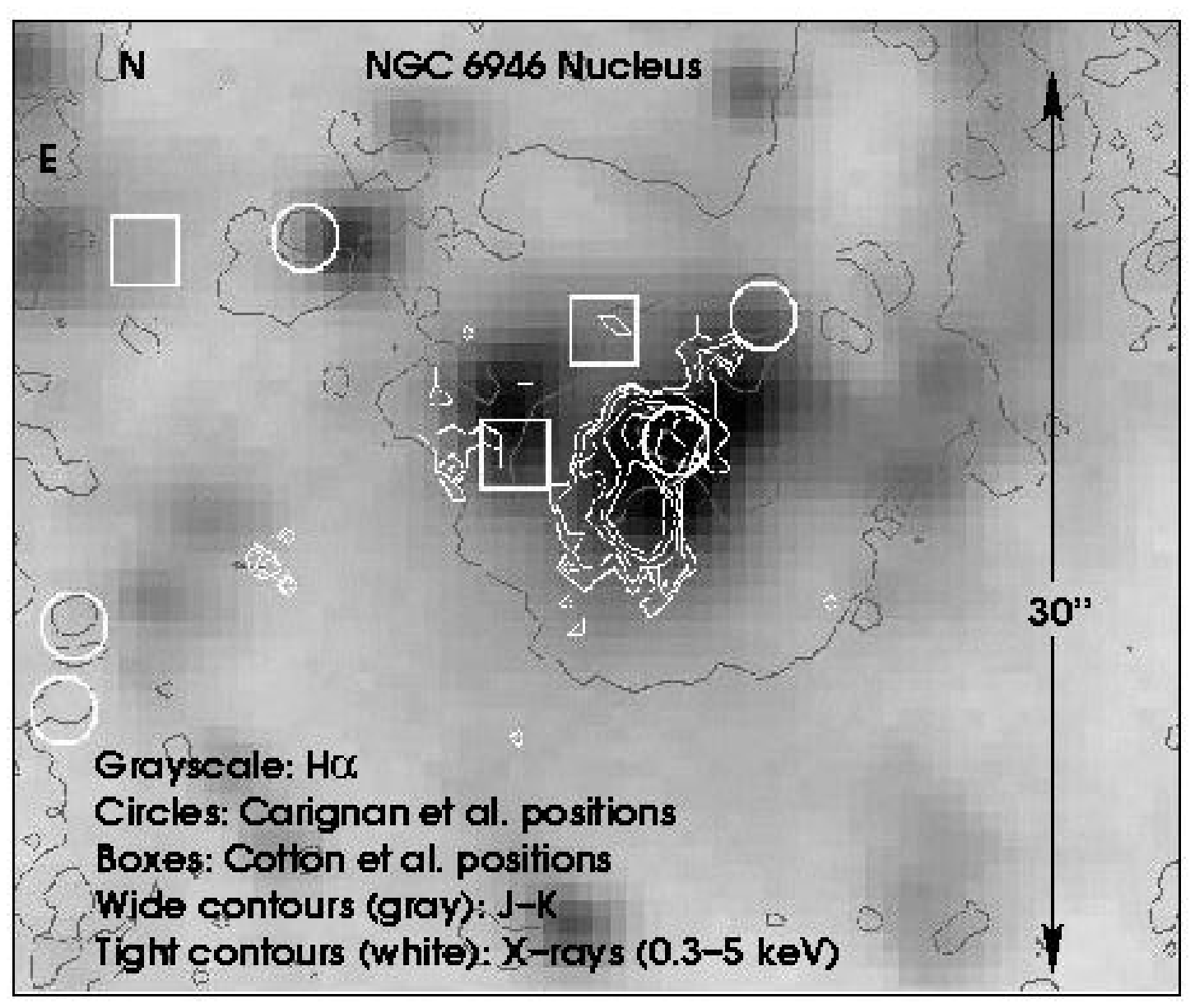}}}
\caption{An expanded view of the nuclear region of NGC 6946.  The
grayscale is the H$\alpha$ image from \citet{MF97} overlaid with
two sets of contours.  The gray, wide contours represent J-K emission
from \cite{Elm98}; the white contours represent 0.3-8 keV X-ray
emission from the nucleus.  The X-ray contours start at 2 counts
pixel$^{-1}$ and reach 20 counts pixel$^{-1}$ in 5 logarithmic steps.
The squares and circles are as in Figure~\ref{nuc_fig}, referring to
the center defined by \cite{Cot99} and \cite{Car90}, respectively.}
\label{nuc_fig_exp}
\end{center}
\end{figure*}

A similar central double source is found in a J$-$K image of the
galactic center of NGC 6946 \citep{Elm98}, but the sources are not
coincident with the two strong central X-ray sources (see
Figure~\ref{nuc_fig} or ~\ref{nuc_fig_exp}).  In fact, there is no
correlation between the 14 J$-$K hot spots and the X-ray sources that
we detect in the central region of the galaxy.  The positions of the
J-K hot spots, presumably associated with H~II regions, are shown
overlaid as circles on the Chandra image in Figure~\ref{nuc_fig}.  The
positions of the H~II regions are inferred from the nuclear position
given in the caption to Figure 11 of \cite{Elm98}, and are plotted
twice: with squares using the Cotton et al. center, and with circles
using the Carignan et al. center.

Although not easily discernible in Figure~\ref{nuc_fig}, but more
easily visible in the contours of Figure~\ref{nuc_fig_exp}, extended
emission is present to the north and to the south; we are certain that
this emission does not originate from the wings of the point spread
function because it does not exhibit the PSF symmetry.  Using the
Carignan et al. position, the contours surround the nuclear H II
region and the extended emission falls along the curve connecting the
northern arm in the J$-$K image to the southern J$-$K extension.
Using the Cotton et al. position, the contours lie W of the J$-$K
center region and do not correspond to any structure in the J$-$K
image.

\section{Discussion}

Luminous ($>$10$^{36}$ erg s$^{-1}$) X-ray sources are most commonly 
degenerate stars accreting mass from non-degenerate companions in 
binary systems.  The degenerate component of these binaries is 
typically a neutron star, since white dwarfs have insufficient surface 
gravity to provide luminous X-ray emission from accretion and stellar 
black holes in binaries appear to be relatively scarce.  The masses of 
the non-degenerate binary components are used to classify X-ray binaries 
as high mass (HMXBs) or low mass (LMXBs), with the boundary in the 
relatively sparsely populated mass region of a few solar masses.

The X-ray LMXB and HMXB populations have now been characterized in
several spiral galaxies, including the Milky Way. LMXBs have a
dynamical lifetime (during which the system is not always
X-ray-luminous) of order 10$^{10}$ years, while the lifetime of HMXBs
cannot exceed the nuclear timescale of the high mass components of the
binary systems, or about 10$^7$ years.  \cite{Grimm02a} have demonstrated
that the HMXB luminosity of spiral galaxies exceeds the luminosity
arising from LMXBs when the star formation rates (SFRs) exceed about
0.005 M$_{\odot}$ per year.  X-ray luminous supernova remnants
primarily arise from the same population of progenitors as HMXBs, but
typically account for no more than a few percent of the HMXB
luminosity \citep{Helf01}.  The timescales suggest that HMXBs and SNRs
should be associated with star-forming regions in spiral arms and
starburst nuclei, while LMXBs should be more uniformly distributed
throughout the galactic bulges and haloes.  

\cite{Grimm02b} have confirmed that the distribution of X-ray binaries
in the Milky Way conform to this general conventional wisdom, i.e.,
that the HMXBs are largely confined to the spiral arms, and that the
LMXBs dominate the bulge (and total source) population, while having a
larger scale height in the galactic plane.  \cite{Grimm02a,Grimm02b}
have also characterized the logN-logS distributions to be expected
from LMXBs and HMXBs.  The logN-logS distribution of HMXBs in both the
Milky Way and in other galaxies of all types seems to have a
``universal'' index of 0.6 (for the integral distribution), with no
evidence of a cutoff. The LMXB slope for the Milky Way is flatter
(index 0.26) at low luminosities, but with a high luminosity cutoff
that will steepen the apparent logN-logS precipitously if the low
luminosity portion of the population is beyond the sensitivity of the
survey.  \cite{Kil02} have suggested that an unbroken logN-logS index 
for starburst galaxies simply reflects the recent birth rate of HMXBs, 
while an aging population of LMXBs should exhibit a unit index break 
to accommodate the removal of sources at a characteristic age from 
an older population.

Recent results show that galaxies with star-forming rates similar
to that of the Milky Way seem to exhibit similar characteristics.
\cite{Ten01} report a steepening of the logN-logS distribution for the 
bulge of M81 for L$_{\rm x}>$4$\times$10$^{37}$ ergs s$^{-1}$ from a
lower luminosity index of 0.5.  A similar 0.5 index seems to
characterize the disk population, but without a corresponding high
luminosity steepening.  It thus appears that the situation is much
like that in the Milky Way, with LMXBs dominating the bulge and with
HMXBs playing a larger role in the star-forming regions of the disk.
The average source spectrum is consistent with a power law with
$\Gamma$=1.6.

M31 seems to demonstrate at least some aspects of the same general
behavior as M81 \citep{Shir01,Tru02,Kaar02}.  The central region of 
the galaxy exhibits the high luminosity steepening characteristic of
domination by LMXBs, while the outer regions of the galaxy exhibit an
index of 1.2.  While this is considerably steeper than 0.6,
\cite{Tru02} assign an origin in faint HMXBs to this population since
the low luminosity index for LMXBs is even less compatible with the 
low luminosity index of LMXBs.  \cite{Shir01} report an average 
source spectrum consistent with a power law with $\Gamma$=1.7.

In contrast, the results reported here for NCG 6946, i.e., for the 
entire galaxy (index 0.64) and for the central 2$'$ containing the
starburst nucleus (0.51), are in good agreement with the expected 0.6
for HMXBs.  The slightly flatter slope for the $>$2$'$ distribution
may reflect a slightly higher contamination by bright LMXBs.  The
global SFR of NGC 6946 is more than an order of magnitude higher than
that of the Milky Way, at approximately 4 M$_{\odot}$ yr$^{-1}$ for 
stars in the mass range 2-60 M$_{\odot}$ \citep{SGC98}; most of this 
is in the outer spiral arms rather than the starburst nucleus.

M83 is a more intense starburst galaxy with a hotter and more compact
photodissociation region than NGC 6946 \citep{IB01}.  \cite{SW02} have
reported that the central region surrounding the starburst nucleus has 
an X-ray logN-logS index of 0.8 (assumed by those authors to arise 
from continuous star formation), while the outer region is somewhat 
flatter (index 0.6) at low luminosities 
($<$6$\times$10$^{37}$ ergs s$^{-1}$)and steeper (index 1.3) at higher 
luminosities, so that most of the high luminosity population of the 
galaxy is in the central region.  This might suggest that in M83 the 
HMXBs dominate the source population only in the central starburst 
region, rather than in the entire galaxy.

M101 is another face-on spiral galaxy which, like the others discussed
above, has a typical H column density through the Milky Way that is as
much as an order of magnitude lower than for NGC 6946.  The global
star formation rate for M101 is very similar to NGC 6946, at 4.8
M$_{\odot}$ yr$^{-1}$, a level about twenty times that of the Milky
Way. The log N-log S distribution reported by \cite{Pen01} is
relatively flat (index of 0.8), and the X-ray sources are clearly
associated with young population stars through their correlation with
the spiral arms.  The distribution of spectral indices for the sources
is broad (per Figure 6 of \citealt{Pen01}), with an average value
similar to that of M81 or M31.  Supersoft sources make up
approximately 10\% of the M101 sample, but this potential similarity
with NGC 6946 cannot be tested, because the higher column density in
the direction of NGC 6946 prevents their detection.

The global results reported here for NGC 6946 are consistent with the
trends discussed above.  With an SFR comparable to that of M101 and
its characterization as a starburst galaxy, it should not be
surprising that the distribution of X-ray sources in NGC 6946 matches
that expected from a population dominated by HMXBs.  The lack of 
obvious blast-wave SNRs may also be expected, in spite of the number 
of recent SNs, for at least two reasons.  From the ratio of their
X-ray-emitting lifetimes, the ratio of X-ray luminous SNRs to HMXBs
originating in the same progenitor sample should be not much more than 
about 1\%.  Further, and most important in this study, a
detection threshold at the level comparable to the luminosity of the 
Crab nebula will be incapable of detecting any but the very brightest 
members of the SNR population of NGC6946.

The galactic center region has two luminous X-ray sources close to the
dynamical center.  Since the luminosity of the southern member of the
center pair is an order of magnitude in excess of that for an
Eddington-limited neutron star, there is the suspicion that it might
be associated with a central black hole.  Although these two sources
seem to fall on the central CO and H alpha images of NGC 6946, the
other X-ray sources near galactic center do not cluster in the CO
molecular bar or correlate in detail with H~II regions identified by
Elmegreen et al. in the central star-forming region of the galaxy.

The other NGC 6946 sources that exceed nominal neutron star Eddington
luminosity can have their explanations in a variety of ways, such as
magnetospheric effects around neutron stars, or black holes in place
of neutron stars (e.g., see \citealt{Grimm02b}).  \cite{Kil02} have
noted that the presence of multiple super-Eddington X-ray sources
seems to be characteristic of starburst galaxies, in marked contrast
to non-starburst spirals.  It is also interesting to note that
stellar-mass black holes in the Milky Way have been identified with
HMXBs that exhibit episodic X-ray increases (e.g., see
\citealt{Kong02}), so that we have local empirical evidence for the
existence of black holes in HMXBs.  A characteristic of such black
hole sources is their spectral variability with luminosity phase; this
latter variation may be the most efficient way to determine the nature
of these sources in subsequent X-ray exposures to NGC 6946.  If they
are stellar black holes in HMXBs, the bright sources detected here are
likely to be in the ``high'' states that usually exhibit spectra
steeper than that of the Crab nebula; in this survey of NCG 6946 the
most luminous sources appear to systematically exhibit spectra steeper
than that of the Crab nebula, at least in the energy range 0.5-5 keV.
Subsequent exposures to NGC 6946 that provide correlated spectral and
intensity variations for these sources could provide the evidence that
they are, indeed, black hole HMXBs.

The previous identification of the ultra-luminous source with the 
supernova remnant MF16 resulted in a plausible explanation for its
high luminosity by \cite{Dun00} in terms of the cooling of a pre-SN 
circumstellar wind that has been recently heated by the SN blast 
wave (although it did not appear that it would be fully adequate 
and these authors suggested a possible additional pulsar component).  
A similar 
scenario had been invoked by \cite{Fab96} to explain the detection 
of X-ray emission from SN1988Z 6.5 years after outburst with the 
{\it Rosat} High Resolution Imager (which has no energy resolution) 
at an inferred X-ray luminosity of $\sim$10$^{41}$ ergs s$^{-1}$. 
The lack of prominent line emission in the current data is difficult 
to reconcile with this model, however.  It is possible to contrive 
scenarios with pre-ionized circumstellar material to suppress line 
emission, but high luminosity arising from bremsstrahlung emission 
at high density will result in rapid cooling to 
temperatures at which X-ray line emission will be prominent.  
The apparent constancy of the luminosity for at least a decade and 
the lack of line emission argue against this being a complete 
explanation.

It may well be that the ultra-luminous source in NGC6946 has its 
ultimate explanation in a very unusual, if not a unique, set of 
circumstances.  One possibility that might appear to have a 
relatively simple set of contrivances is a short-period pulsar with 
a high surface magnetic field.  The 0.5-5 keV luminosity of the 
Crab nebula, with a central pulsar having a rotation period of 
33 ms and an inferred surface magnetic field of the order of 
10$^{12}$ gauss, is approximately three orders of magnitude lower 
than that of MF16.  Since the loss of pulsar 
rotational kinetic energy is proportional to the square of the 
surface field and the inverse fourth power of the period, a factor 
of three in both of these parameters relative to those of the 
Crab pulsar could provide a factor of 10$^{3}$ in luminosity.  
There are some obvious difficulties with this simple picture, 
however.  There are no reported high-field pulsars with short 
periods or short-period spin-down pulsars with high fields 
(the fastest period for a pulsar in a remnant is 16 ms in N157B, 
with an inferred magnetic field of order 10$^{8}$ gauss, 
\citealt{Mars98}).  The lack of a Crab-like gamma-ray source at 
the position of NGC 6946 \citep{Har99} is not particularly 
troublesome, since the distance to NGC 6946 would reduce the 
simple scaling of a source with gamma-ray luminosity 10$^{3}$ that 
of the Crab nebula to an apprarent luminosity of only 10$^{-4}$.  
A more serious difficulty with this explanation is that it demands 
a very young pulsar; we cannot sustain a very short pulsar period 
with a high luminosity that is the direct consequence of pulsar 
rotational energy loss.  The implied characterisitic age 
for such a pulsar would be about 150 years, which is inconsistent 
with optical loop sizes of tens of pc and optical shock velocities 
of 600 km/s.  We might reinvoke multiple SN to explain the 
optical characteristics of the region into which this anomalously 
bright pulsar was recently introduced, but the same coincidence 
could just have easily provided the setting for a HMXB with a 
massive stellar black hole.  Future observations of the secular 
variation of the source intensity and/or spectrum should provide 
additional clues to the nature of this source.

\section{Summary}

We have utilized the {\it Chandra} ACIS imaging spectrometer to 
investigate the source population of the face-on galaxy NGC 6946.
The gross aspects of the spatial and luminosity distributions of 
the source population suggest that HMXBs may play a significantly 
more important role here (relative to LMXBs) than in the 
Milky Way, consistent with the much higher SFR in NGC 6946.  
Although there are strong nuclear sources, including one that may 
be coincident with the center of the galaxy, there is no 
clustering of this sample in the CO molecular bar or correlation 
of the central X-ray sources with near IR ``hot spots.''  The ten 
brightest sources in the galaxy appear to exceed the Eddington 
luminosity for a neutron star, and the very brightest, previously 
identified with a supernova remnant (MF16),
exceeds it by two orders of magnitude.  The most straightforward 
conjecture for the nature of all of these strong sources is that 
they are HMXBs with black hole degenerate members.  In the case 
of MF16, the lack of strong X-ray emission lines in its X-ray 
spectrum suggests that the major fraction of its luminosity 
cannot be reconciled with traditional collisional models for 
the X-ray emission from supernova remnants.

Thanks to Olin undergraduate students Polina Segalova, Jessica 
Anderson, Anne Marie Rynning and Nicholas Zola for help with 
data extraction and analysis during the early phases of this 
study.  The research of EMS was supported by contract 
number NAS8-38073 to SAO for the Chandra X-ray Center.

\clearpage


\begin{table*}
\scriptsize
\begin{center}
\caption{Detected Sources, Hardness Ratios, Identifications}
\label{detsrcs}
\begin{tabular}{rllrrrrl}
        & \multicolumn{2}{c}{J2000} & L$_{\rm X}$ & Source  & \multicolumn{2}{c}{Identities} \\
 No. & RA & Dec & 0.5-5 keV & Hardness  & SBF\tablenotemark{a} & LDG\tablenotemark{b} & Comment \\ \hline
 1 & 308.5357 & +60.1712 & 10   & & \\
 2 & 308.5792 & +60.1689 & 62   & M & & 1 \\
 3 & 308.5890 & +60.1716 & 30   & & \\
 4 & 308.6029 & +60.1774 & 48   & H & \\
 5 & 308.6039 & +60.1517 & 170  & M & \\
 6 & 308.6088 & +60.1527 & 700  & M & S1 & \\
 7 & 308.6093 & +60.1543 & 40   & & \\
 8 & 308.6212 & +60.1808 & 26   & & \\
 9 & 308.6313 & +60.1106 & 11   & & \\
10 & 308.6352 & +60.1393 & 50   & H & \\
11 & 308.6434 & +60.1758 & 106  & M/S & S2 & \\
12 & 308.6470 & +60.1285 & 16   & & \\
13 & 308.6503 & +60.1443 & 42   & S & & 17 \\
14 & 308.6515 & +60.0995 & 33   & & & 18 \\
15 & 308.6520 & +60.1584 & 810  & M/S & S3 & \\
16 & 308.6549 & +60.1650 & 11   & & \\
17 & 308.6555 & +60.1718 & 11   & & \\
18 & 308.6576 & +60.0760 & 36   & & \\
19 & 308.6657 & +60.1397 & 59   & H & & 24 \\
20 & 308.6697 & +60.1407 & 10	\\
21 & 308.6719 & +60.1461 & 20	& & & 26 \\
22 & 308.6752 & +60.0915 & 12   \\
23 & 308.6801 & +60.1144 & 13   \\
24 & 308.6901 & +60.1560 & 11   \\
25 & 308.6916 & +60.1804 & 23   \\
26 & 308.6930 & +60.2246 & 49   & H/M \\
27 & 308.6962 & +60.1474 & 29   \\
28 & 308.6980 & +60.1675 & 34   & & & 33 \\
29 & 308.6983 & +60.1592 & 12   \\
30 & 308.7018 & +60.1365 & 185  & M \\		
31 & 308.7028 & +60.1616 & 25   \\
32 & 308.7032 & +60.0984 & 335  & S & S4 \\
33 & 308.7043 & +60.2049 & 210  & M \\
34 & 308.7064 & +60.2286 & 24   \\
35 & 308.7071 & +60.1699 & 20   \\
36 & 308.7072 & +60.1699 & 25   \\
\end{tabular}
\tablenotetext{a}{sources from \cite{SBF00}.}
\tablenotetext{b}{sources from \cite{LDG97}.}
\end{center}
\end{table*}


\clearpage

\begin{table*}
\tablenum{1 (continued)}
\scriptsize{\footnotesize\tiny}
\begin{center}
\caption{Detected Sources, Hardness Ratios, Identifications}
\label{detsrcs1a}
\begin{tabular}{rllrrrrl}
37 & 308.7076 & +60.1349 & 13   & & & 43 \\
38 & 308.7077 & +60.1721 & 10   \\
39 & 308.7085 & +60.1568 & 41   & M \\
40 & 308.7104 & +60.1710 & 10   \\
41 & 308.7116 & +60.2019 & 11   \\
42 & 308.7120 & +60.1294 & 15   \\
43 & 308.7121 & +60.1725 & 66   & S & & 48 \\
44 & 308.7149 & +60.2139 & 12   \\
45 & 308.7180 & +60.1533 & 780  & M & S5 & & nucleus \\
46 & 308.7180 & +60.1540 & 135  & M & & & nucleus \\
47 & 308.7224 & +60.1221 & 10   \\
48 & 308.7225 & +60.1819 & 18   & & & 67 \\
49 & 308.7318 & +60.1584 & 10   \\
50 & 308.7354 & +60.1428 & 380  & H/M & S6 \\
51 & 308.7356 & +60.1387 & 25   & & & 79 \\
52 & 308.7367 & +60.0923 & 46   & H/M \\
53 & 308.7405 & +60.1636 & 340  & M & S7 \\
54 & 308.7434 & +60.1595 & 38   & & & 82 & 1968D \\
55 & 308.7440 & +60.2382 & 84   & H/M \\
56 & 308.7505 & +60.1522 & 1800 & S & S9 \\
57 & 308.7527 & +60.2139 & 43   & H \\
58 & 308.7531 & +60.1919 & 8400 & M/S & S10/8 & 85 & MF16\\
59 & 308.7540 & +60.1697 & 380  & H/M & S11 \\
60 & 308.7581 & +60.1676 & 170  & H/M \\
61 & 308.7692 & +60.1882 & 73   & M/S & & 91 \\
62 & 308.7697 & +60.1551 & 15   \\
63 & 308.7722 & +60.1064 & 90   & H/M \\
64 & 308.7770 & +60.1781 & 15   \\
65 & 308.7830 & +60.1871 & 15	& & & 101 \\
66 & 308.8033 & +60.1254 & 191  & H/M & S12 \\
67 & 308.8177 & +60.1796 & 51   & H \\
68 & 308.8284 & +60.1823 & 335  & M & S13 \\
69 & 308.8335 & +60.1597 & 15   \\
70 & 308.8380 & +60.1644 & 25   \\
71 & 308.8561 & +60.1665 & 78   & S & S14? \\
72 & 308.8661 & +60.1553 & 36   \\ \hline	
\end{tabular}
\tablenotetext{a}{sources from \cite{SBF00}.}
\tablenotetext{b}{sources from \cite{LDG97}.}
\end{center}
\end{table*}

\clearpage

\begin{table*}
\begin{center}
\scriptsize
\caption{Spectral Fit Results}
\label{spec_mod}
\begin{tabular}{rlllllll}
\hline\hline
Src ID\tablenotemark{a} & Box  & Model\tablenotemark{b}  & ${\chi}^2/dof$, 
$dof$ & N$_{\rm H}$     &${\Gamma}$ or kT & \multicolumn{2}{c}{Flux 
(10$^{-13}$ erg s$^{-1}$ cm$^{-2}$)} \\
   & (pixels\tablenotemark{d}~) & & & ($\times 
10^{22}$) &(- or keV) & Observed\tablenotemark{c}  & Source \\ \hline
6 (RA=308.609) &8$\times$10 &PL &1.36, 67 &0.34 (0.27-0.44) &2.1 (1.9-2.3) 
&0.84 &1.3 \\
           & &Br &1.26, 67 &0.25 (0.18-0.33) &3.6 (2.8-5.0) &0.87 
&1.1 \\ \\
15 (RA=308.652)&4$\times$6 &PL &1.22, 51 &0.44 (0.33-0.51) &4.5 (4.0-5.0) 
&0.50 &2.7 \\
           & &Br &1.35, 51 &0.16 (0.10-0.22) &0.69 (0.56-0.82)&0.50 
&0.81\\ \\
32 (RA=308.703)&10$\times$10&PL&1.82, 24 &0.59 (0.37-0.85) &6.6 (5.2-8.3) 
&0.19 &4.0\\
           & &Br &1.94, 24 &0.32 (0.13-0.47) &0.27 (0.20-0.43) & 0.18 & 0.78\\
           & &PSH$^e$&1.37, 23 &0.57 (0.52-0.69) &1.0 
(0.51-1.9)&0.21 &3.3\\ \\

45 (RA=308.718)&$12\times$12&PL&1.07, 85 &0.60 (0.51-0.66) &2.6 (2.4-2.8) &1.1 
&2.8\\
           & &Br &1.05, 85 &0.41 (0.36-0.47) &2.4 (2.0-2.8) &1.2 
&1.9\\ \\

50 (RA=308.735)&$6\times$6  &PL &1.63, 28 &0.78 (0.55-1.22) &1.78 (1.67-2.25) 
&0.43 &0.77 \\
           & &Br &1.60, 28 &0.63 (0.44-0.94) &7.8 (4.0-16) &0.42 
&0.66\\
           & &PSH$^f$&1.48, 27 &0.83 (0.53-1.04) &5.9 (3.4-10.) 
&0.44 &3.6 \\ \\
53 (RA=308.741)&8$\times$10 &PL &1.28, 28 &0.64 (0.53-0.83) &2.8 (2.6-3.0) 
&0.32 &0.90 \\
           & &Br &1.27, 28 &0.45 (0.32-0.59) &1.9 (1.6-2.5) &0.32 
&0.58 \\ \\
56 (RA=308.751)&10$\times$12&PL &1.58, 85 &0.52 (0.45-0.59) &4.9 (4.6-5.3) 
&1.2 &9.3\\
           & &Br &1.84, 85 &0.24 (0.19-0.29) &0.53 (0.46-0.61) &1.1 
&2.5 \\ \\
58 (RA=308.753)&14$\times$20&PL &1.49, 204 &0.15 (0.13-0.17)&2.44 (2.36-2.51) 
&7.1 &9.8 \\
           & &Br &1.90, 204 &0.031 (0.019-0.043) &2.5(2.3-2.7) &7.1 
&7.5 \\
           & &2Br&1.17, 202 &0.42 (0.34-0.48) &3.2 (2.9-3.7) &4.7 
&7.4\\
           & & & & &0.23 (0.20-0.27) 
&2.4 & 16.5\\
           & &MKL+PL&1.11, 202 &0.12 (0.10-0.14) &kT=0.78 
(0.63-0.84) &0.65& 0.91 \\
           & & & & &$\Gamma$=2.2 
(2.1-2.3) &6.3 & 8.0 \\ \\
59 (RA=308.754)&10$\times$14&PL &1.10, 33  &0.30 (0.21-0.50) &1.7 (1.5-2.0) 
&0.42 &0.57 \\
           & &Br &1.04, 33  &0.27 (0.16-0.39) &6.8 (4.1-13.) &0.41 
&0.53 \\ \\
68 (RA=308.828)&24$\times$24          &PL &1.08, 38  &1.2 (1.0-1.5)    &3.1 
(3.0-3.5) &0.52 &2.6 \\
           & &Br &0.86, 38  &0.86 (0.70-1.07) &1.8 (1.4-2.1) &0.55 
&1.3 \\ \hline
\end{tabular}
\tablenotetext{a}{Source number from Table 1.}
\tablenotetext{b}{Models: PL=power-law; Br=bremsstrahlung; MKL=MEKAL; 
NEI=nonequilibrium ionization; PSH=plane-parallel shock}
\tablenotetext{c}{Detected 0.5$-$5.0 keV flux}
\tablenotetext{d}{1 ACIS pixel is 0.5$''$.}
\tablenotetext{e}{Model sums ionization ages from 0 to $n_et = 1.1 (0.84-1.6) 
\times 10^{10}$ cm$^{-3}$ s.}
\tablenotetext{d}{Model sums ionization ages from 0 to $n_et = 2.5 (1.2-3.3)  
\times 10^{9}$ cm$^{-3}$ s.}
\end{center}
\end{table*}

\clearpage

\clearpage

\clearpage

\clearpage

\clearpage

\end{document}